# Ultra low energy results and their impact to dark matter and low energy neutrino physics

E. Bougamont[1], P. Colas[1], J. Derre[1], I. Giomataris[1], G. Gerbier[1], M. Gros[1], P. Magnier[1], X.F. Navick[1], P. Salin[3], I. Savvidis[2], G. Tsiledakis[1], J. D. Vergados[4]

*1 : IRFU, Centre d'études de Saclay, 91191 Gif sur Yvette CEDEX, France*
*2 : Aristotle University of Thessaloniki, Greece*
*3 : APC, Université Paris 7 Denis Diderot, Paris, France*
*4: University of Ioannina, Greece*

**Abstract**

We present ultra low energy results taken with the novel Spherical Proportional Counter. The energy threshold has been pushed down to about 25 eV and single electrons are clearly collected and detected. To reach such performance low energy calibration systems have been successfully developed:
- A pulsed UV lamp extracting photoelectrons from the inner surface of the detector
- Various radioactive sources allowing low energy peaks through fluorescence processes.

The bench mark result is the observation of a well resolved peak at 270 eV due to carbon fluorescence which is unique performance for such large-massive detector. It opens a new window in dark matter and low energy neutrino search and may allow detection of neutrinos from a nuclear reactor or from supernova via neutrino-nucleus elastic scattering

**I. Introduction**

The development of massive low-background, low-energy threshold detectors is a challenge in contemporary physics.

The search of Dark matter in form of hypothetical Weakly Interacting Massive Particles (WIMP) is under intense development and relies on the detection of low energy (keV scale) recoil produced by their elastic nuclear interaction with detector nuclei. This recoil is then detected, either via ionization, scintillation or phonon signatures. For that several detectors have been developed [1].

The virtue of detectors able to reach sub-keV energy threshold was recognized by many authors for studying dark matter at lower energy than present experiments or addressing low energy neutrino physics [2,3].

Current experimental searches are also developed with regard to the claimed detection of a signal by DAMA group and also some excess observed at lower energy.
The DAMA collaboration reported an observation of an excess in the detection rate in the low energy spectrum together at 3 keV which exhibits seasonal variation [4]. The need to go to very low energies may become more crucial, if the WIMPs turn out to be very light, since, then, the energy transfer to the nucleus is smaller. This is also true in the case of the recently discussed secluded dark matter, since, in this case, the interaction with quarks is mediated by the massless photon [5].

New generation of directional detection of dark matter is under development to perhaps allow for an unambiguous observation even in presence of backgrounds and to observe directional anisotropy of the recoils [6,7,8].

The scattering of low-energy neutrinos off nuclei (few MeV for nuclear reactor emission) via the neutral current remains undetected thirty five years after its first description [9,10], a great challenge in neutrino physics. Experiments with reactor neutrinos will provide a first step in verifying coherent neutrino scattering. If the coherence effect occurs, the resulting cross section is very large compared to other low-energy neutrino interaction channels. Because the neutrino is light the recoil of nuclei is extremely small and the challenge is at the low threshold required (typically below 100 eV).

As it has already been mentioned at low neutrino energies the neutrino-nucleus neutral current interaction becomes very important. The reason is that all neutrons in the nucleus contribute coherently. The detection consists in observing the recoiling nucleus. The energy transferred to the nucleus is quite small and decreases with the mass of the target. The recoil energy depends on the scattering angle, its optimum is achieved in the forward direction. If both the neutrino energy and the recoil energy are measured in units of the recoiling mass one obtains a universal plot (see Fig. 1).

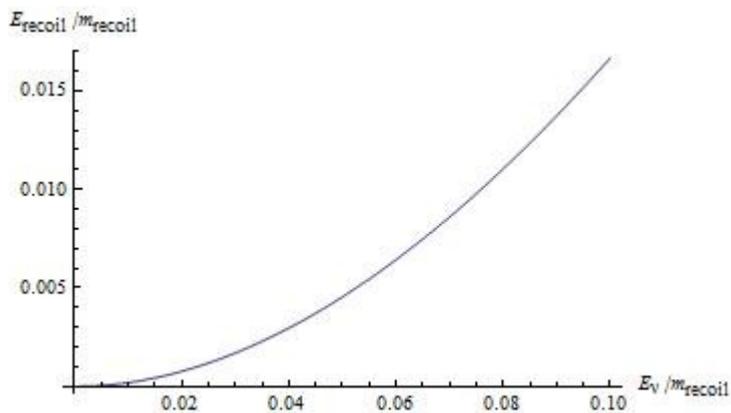

Figure 1: The maximum recoiling energy versus the neutrino energy (both in units of the recoiling mass).

In the case of nuclear recoils we get the picture of Fig. 2.

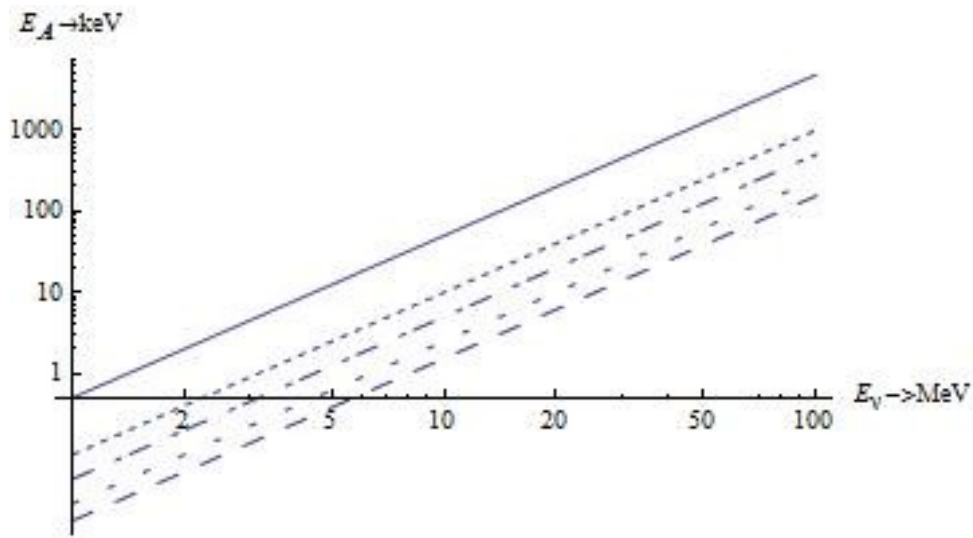

Figure 2: The nuclear recoil energy versus the neutrino energy. From top to bottom nuclear

targets with A=4, 20, 40, 84, 131 for the elements He, Ne, Ar, Kr and Xe respectively.

We will now consider some cases of special interest:
1) Supernova Neutrinos. In this case the emitted neutrinos have a spectrum, which for each flavor is characterized by a Fermi-Dirac distribution with essentially two parameters a temperature and a chemical potential [11]. The peak energies are approximately 15, 25 and 35 MeV for electron neutrinos, electron antineutrinos and all other flavors respectively. The distribution becomes negligible for energies above 70 MeV. Similar phenomenological spectra can be found in the literature. For a detector of radius 4 m with a gas under 10 Atm and a typical supernova in our galaxy, i.e. 10 kpc away, one finds 1, 30, 150, 600 and 1900 events for He, Ne, Ar, Kr and Xe respectively [11]. The effect of threshold is shown in Fig. 3.

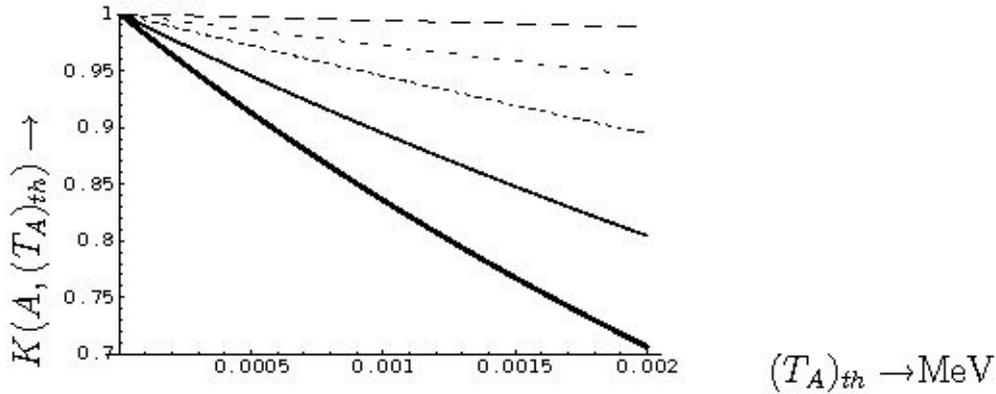

Figure 3: The ratio of the total rate with a given energy threshold divided by that with zero threshold as a function of the threshold energy in MeV. From top to bottom for He, Ne, Ar, Kr and Xe respectively.

2) The Oak Ridge neutron spallation source [12]. This has the great advantage that the neutrino spectrum is well understood. In this instance one has i) an electron neutrino spectrum with a peak energy at 35 MeV and a maximum at 50 MeV, ii) a muon antineutrino spectrum that extends from zero to 50 MeV where it falls abruptly and iii) a discreet muon neutrino with an energy of 30 MeV. Thus this source may be the best place to test the coherent neutrino nucleus cross section mediated by the neutral current. The total rate for a spherical detector filled with Xe and placed 50 m away from the source, assuming zero threshold is 18 events per ton per year. The effect of threshold is shown in the case of the electron neutrino for a Xe target in Fig. 4.

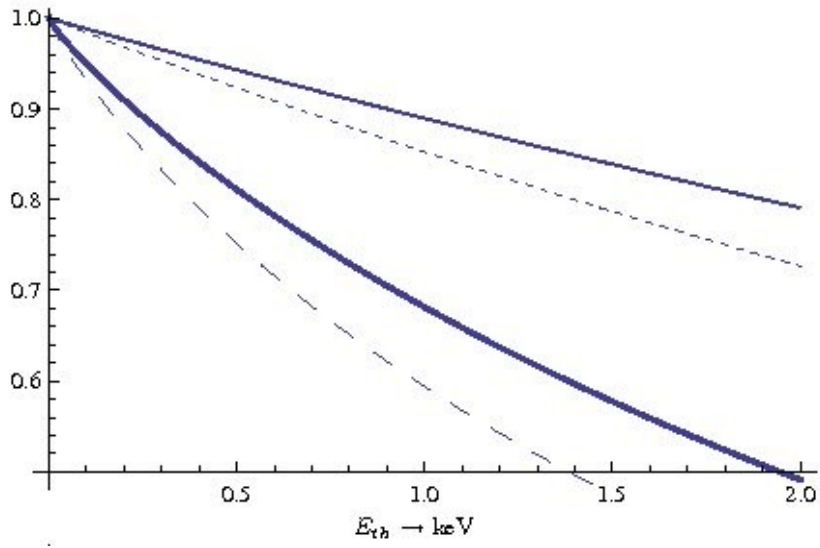

Figure 4: The ratio of the total rate with a given energy threshold divided by that with zero threshold as a function of the threshold energy for a Xe target. The solid (dashed) curves correspond to no nuclear form factor (nuclear form factor) respectively. The two lower curves take into account the effect of quenching.

3) Reactor Neutrinos. The spectrum of these neutrinos depends on the reactor and its mode of operation. This is known quite well for some reactors, e.g. Bugey. A reasonable approximation to a typical reactor spectrum can be found in the literature [13]. A normalized version is given in Fig. 5.

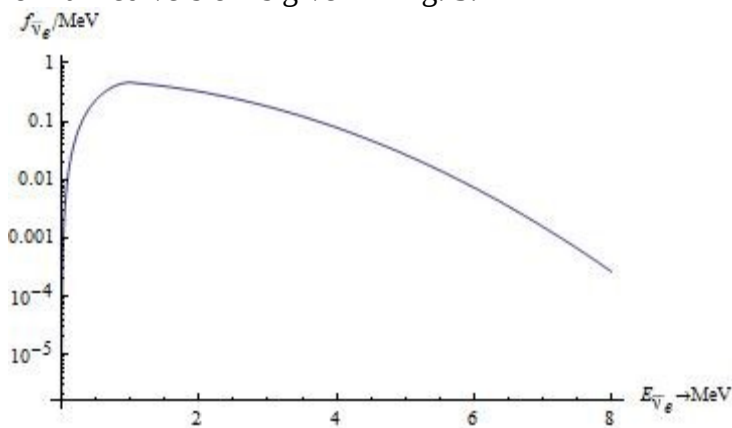

Figure 5: The typical reactor antineutrino spectrum.

We see that the peak energy (~1MeV) and the average energy ( ~2.1 MeV) are quite low. Therefore a neutral current detector is going to be useful, if a low threshold (less than 0.1 keV) and high resolution are achieved. Under these conditions such neutrinos can be used to check the coherent neutrino nucleus scattering.

4) Geoneutrinos. Neutrinos offer a good prospect for an accurate and detailed study of the Earth's interior [14]. Neutrinos of this origin have an energy distribution with a maximum of around 3 MeV and an average of 3.3 MeV. The actual shape is quite complicated [15, 16], but for our purposes a reasonable analytic approximation can be found (Fig.6). Up to now the detectors utilize the interaction of electronic antineutrinos with a proton to produce via charged current exchange a positron and a neutron. The threshold energy is 1.4 MeV as in the future LENA detector [17]. With such high threshold, one can contemplate that a nuclear recoil experiment employing a much lighter nuclear target with threshold in the 0.1 keV region may dare compete with a detector of LENA's caliber. Admittedly the charged reaction has the advantage that by observing both the positron and the neutron one may have directionality information. It is however possible that this can be achieved by directional experiments in which the direction of recoiling nucleus is also observed. The same thing can be achieved by a network of standard recoil detectors, which as we have already mentioned can be robust and cheap.

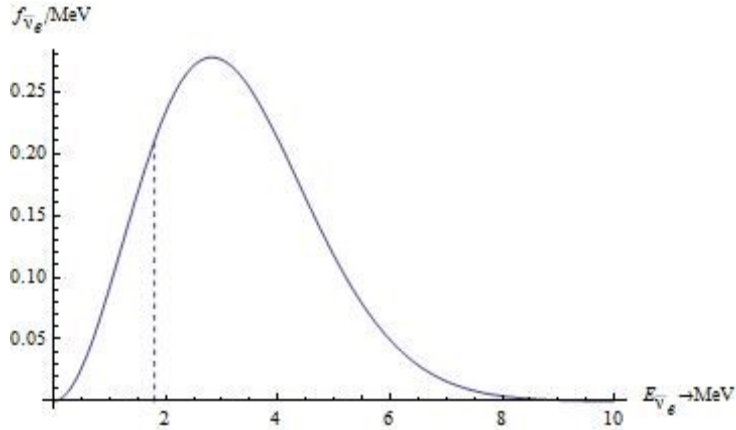

Figure 6: The expected normalized geoneutrino spectrum. The dashed line indicates a typical threshold in other experiments.

Ultra low-noise germanium detectors with sub-keV energy capability have been recently developed and are operating at underground laboratories [18,19]. A small excess observed around 1 keV needs to be clarified and verified by detectors having a lower energy threshold.

In this paper we will report results at low energy obtained by using the novel Spherical Proportional Counter (SPC) which has been recently developed.

**II. Detector description**

The detector consists of a large spherical copper vessel 1.3 m in diameter and a small metallic ball 16 mm in diameter located at the center of the drift vessel, which is the proportional counter. The ball is maintained in the center of the sphere by a stainless steel rod and is set at high voltage. A second electrode (umbrella-shaped) that is placed 24 mm away from the ball, is powered with an independent but lower high voltage, serving as electric field corrector. The detector operates in a seal mode: the spherical vessel is first pumped out and then filled with an appropriate gas at a pressure from few tens of mbar up to 5 bar. Electrons originating from ionization of the gas in the volume drift to the region of the central ball where the intense electric field allows gas amplification to occur. The produced signal is amplified through a charge amplifier and a shaper and is read-out by a 14 bit ADC. Detailed description of the detector, its electronics, its operation and its performance could be found in references [20,21,22].

In a previous paper [22] we reported results obtained at higher energy and we pointed out the excellent energy resolution obtained with radon nuclide and its daughters. In this work we will focus our studies to detect very-low energy gamma or *X*-rays emitted by radioactive sources or fluorescence process.

**III. Low energy calibration and results**

**1. Results with radioactive sources and fluorescence *X*-rays**

For this study we are using a gas filling of Argon with 2% admixture of $CH_4$ at various pressures. The pressure (*P* in mb), the high voltage of the ball (*HV1* in V), the high voltage of the umbrella field corrector (*HV2* in V) and the gain (*G*) of the amplifier-shaper in all measurements in the section, are written in the top-right box on each of the following figures. Calibration of the counter was initially performed using a $^{109}$Cd source by irradiating the gaseous volume through a thin 200 μm aluminum window.

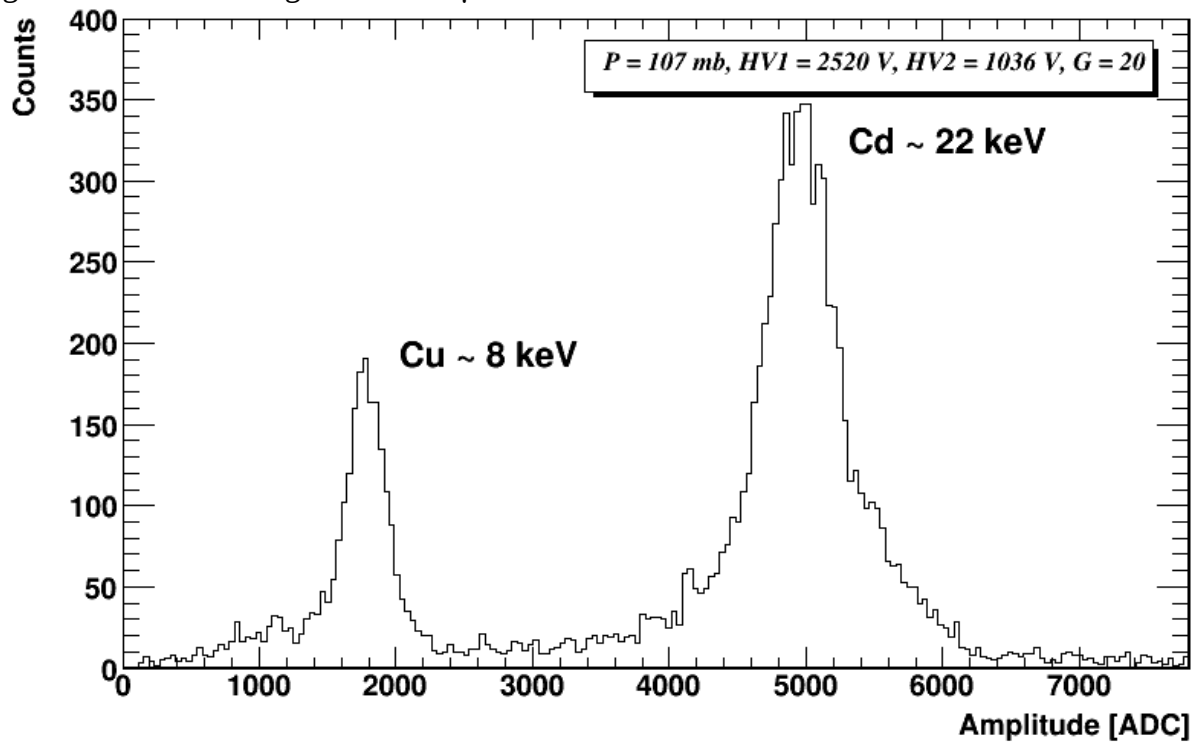

Figure 7: The spectrum of the $^{109}$Cd source with the energy lines of 8 and 22 keV respectively.

Fig. 7 shows the energy spectrum with the 22 keV line from the $^{109}$Cd source and the 8 keV line, which is an induced fluorescence at the copper vessel. The energy resolution is quite satisfactory 6% and 9% (FWHM) respectively.

Gamma fluorescence is adequate for producing fluorescence lines in the range above several keV. However, it is very difficult to produce low energy calibration lines below a few keV. In order to create lower energy *X*-rays we have used an $^{241}$Am source, which decays by the following process: $^{241}$Am → ($^{237}$Np)$^*$ + $^4$He + 5.6 MeV. The $^{237}$Np nucleus then decays into a lower energy state by emitting a 59.537 keV gamma ray and other L rays, which are used to fluoresce the elements. The source was evaporated to a stainless steel holder and attached to the sensor rod at middle distance; the source then is covered by thin foils with adequate thickness to totally absorb the 5.6 MeV alpha emitted, leaving only gamma rays and fluorescence induced *X*-rays to pass into the gas volume.

Covering the source with a 20 μm thick aluminum foil we were able to fluoresce the K *X*-rays ranging from Aluminum whose K electron has a binding energy of 1.56 keV, to Cu whose K electron has a binding energy of 8.98 keV.

Fig. 8 shows the obtained low energy spectrum. From left to right we observe the aluminum line (1.45 keV), the iron line (6.4 keV), the copper line (8.0 keV) and two lines emitted by $^{237}N_p$ nucleus at 13.93 keV (*X*-ray Lα) and 17.61 keV (*X*-ray Lβ).

By increasing the gain of the amplifier 5 times and adjusting its settings, we were able to push the higher part of the spectrum outside the ADC range and keep in the ADC acceptance only the aluminum 1.45 keV peak as shown in Fig. 9. The energy threshold is clearly below 100 eV and the low energy background level is flat.

In order to obtain even lower energy calibration lines we replaced the aluminum foil by a thinner 10 μm one and we attached a 20 μm polypropylene foil. Therefore, the alpha particle that is crossing the aluminum foil, is fully absorbed by the polypropylene foil and induces both aluminum and carbon fluorescence as shown in Fig. 10.

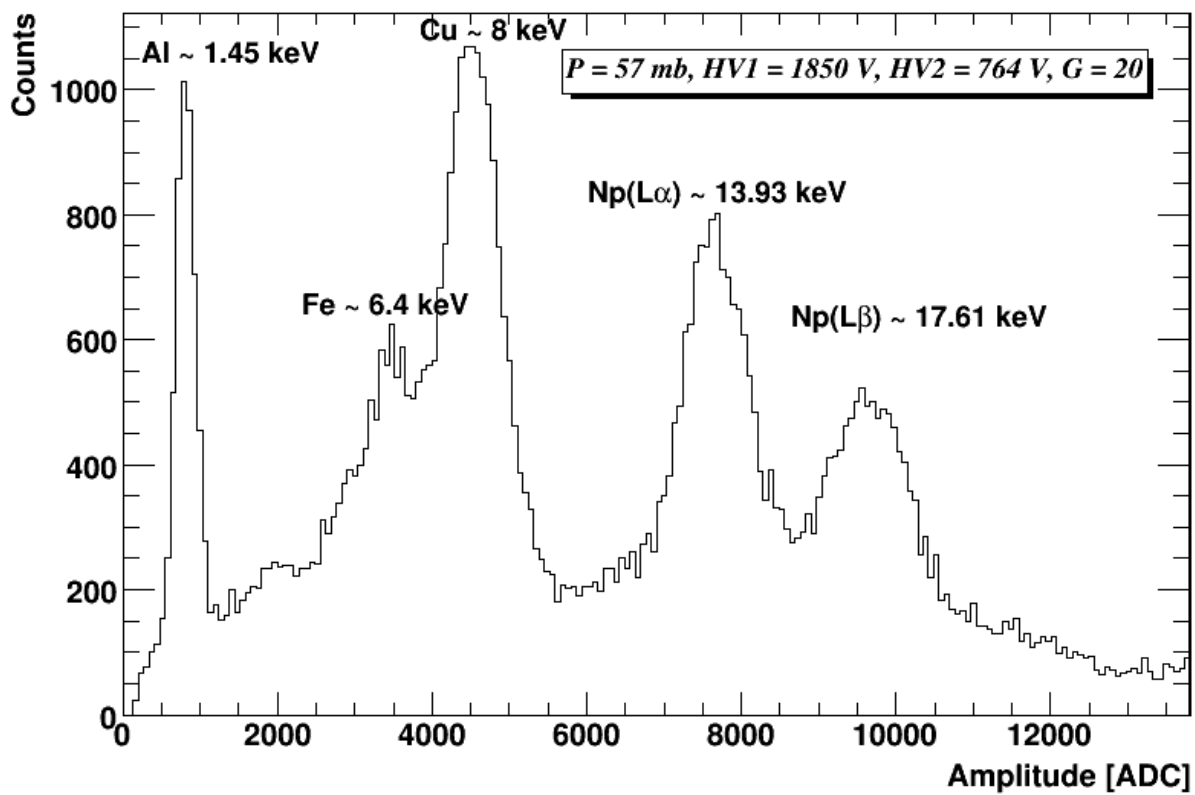

Figure 8: Peaks observed from the $^{241}$Am radioactive source. From left to right we observe the Aluminium, Iron and Copper peaks followed by the Neptunium peaks.

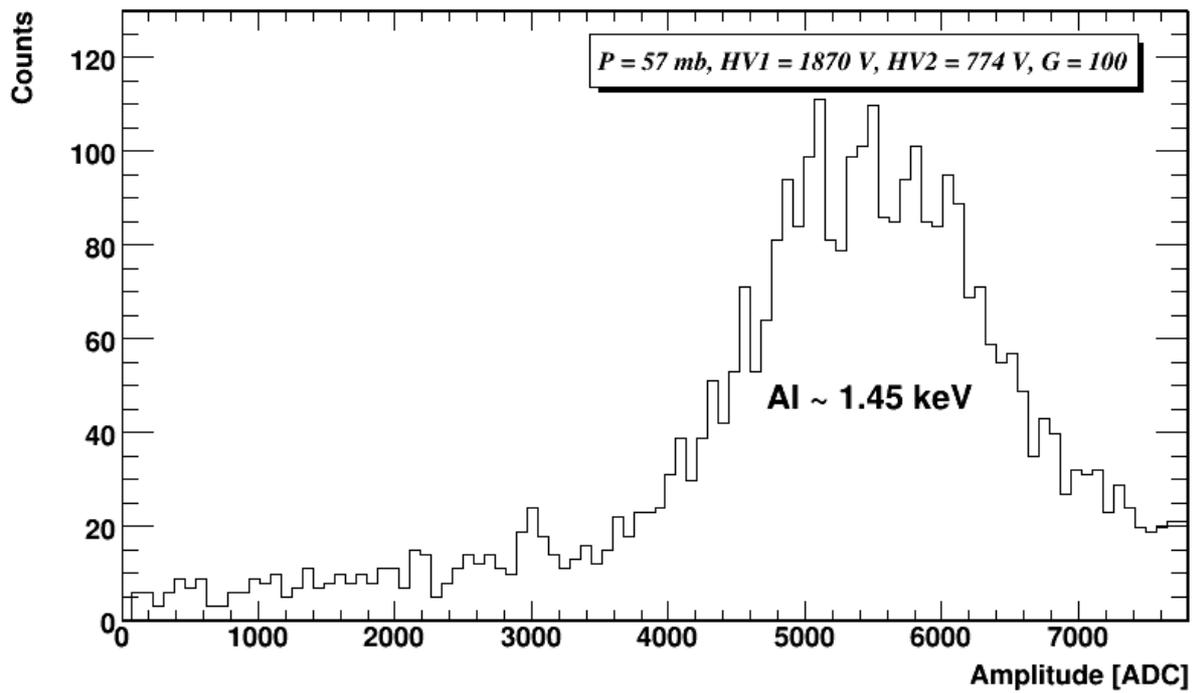

Figure 9: An *X*-ray spectrum showing its characteristic Aluminum K *X*-ray energy peak.

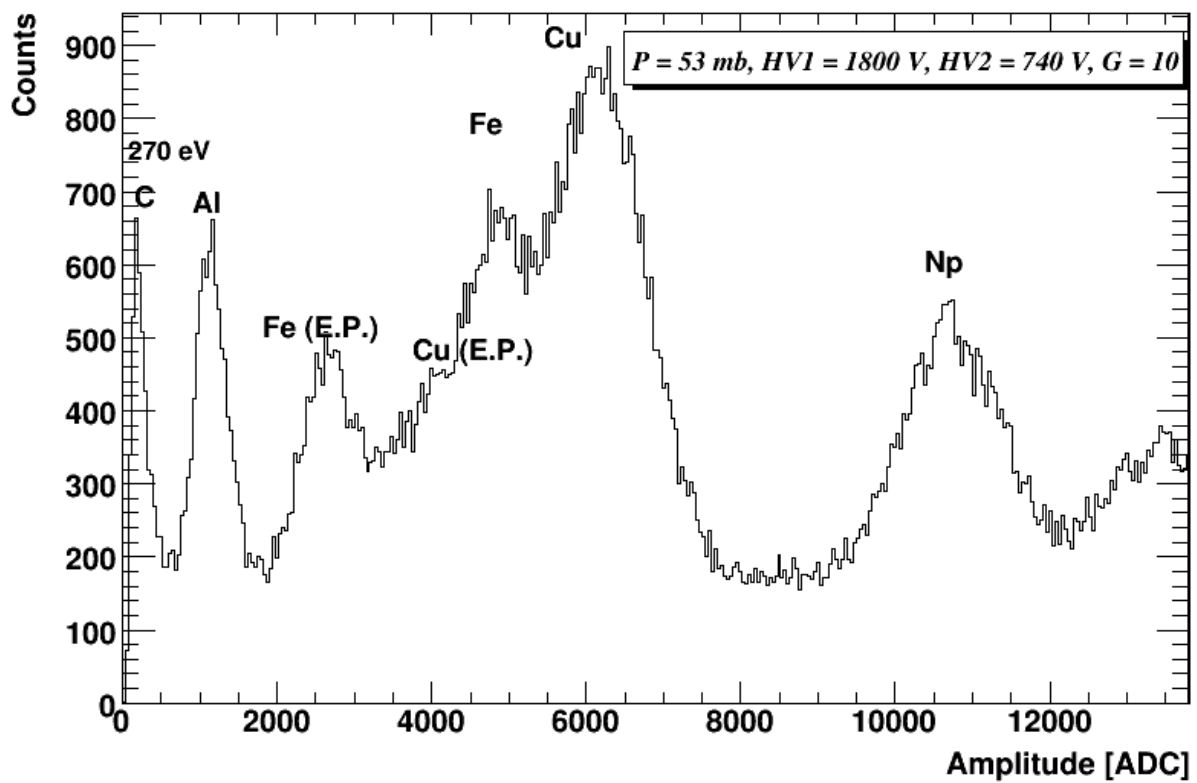

Figure 10: Peaks observed from the $^{241}$Am radioactive source through aluminium and polypropylene foil. On the left the Carbon (270 eV) peak is shown, followed by the Aluminium peak (1.45 keV), the escape peak (E.P.) of Iron in Argon (3.3 keV), the escape

peak of Copper in Argon (5 keV), the Iron peak (6.4 keV), the Copper peak (8 keV) and the Neptunium peak (13.93 keV) .

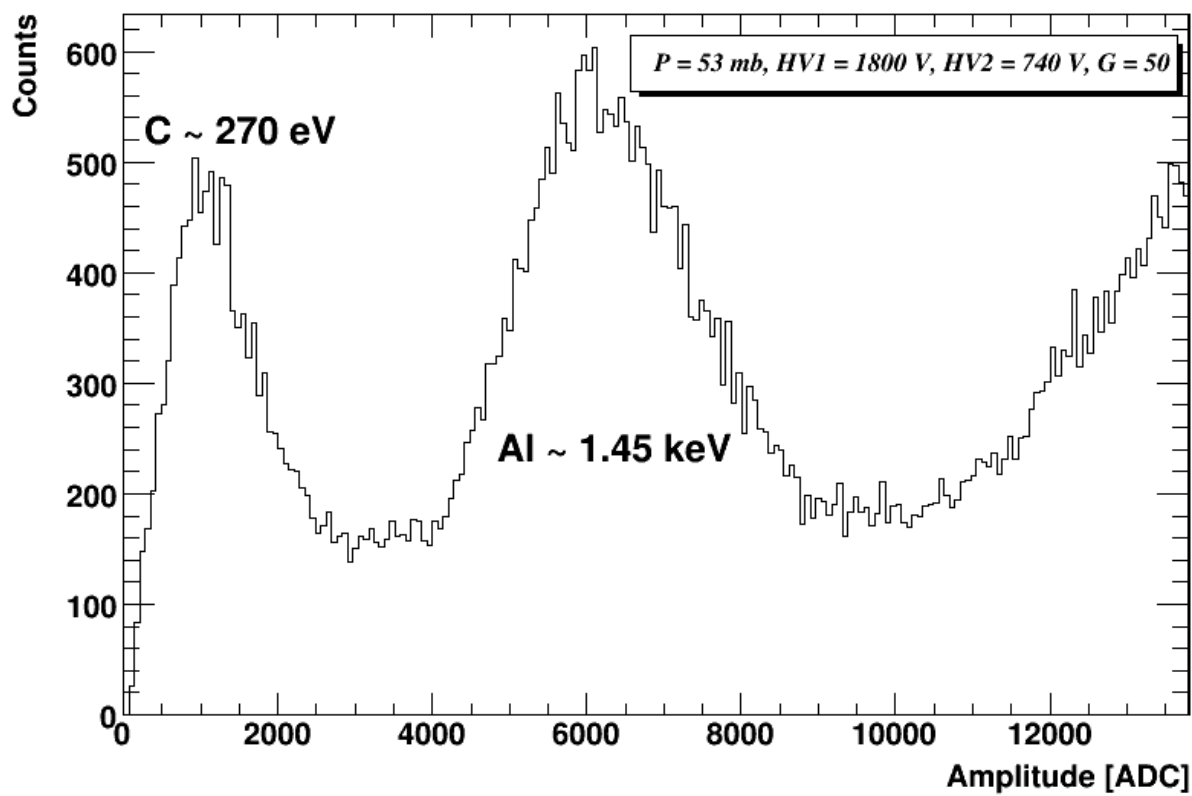

Figure 11: From left the Carbon (270 eV) peak followed by the Aluminum peak (1.45 keV).

By increasing the gain of the amplifier by a factor of 5, we are able to push the higher part of the spectrum outside the ADC range and keep in the ADC acceptance only the Carbon 270 eV peak as well as the Aluminum (1.45 keV) as shown in Fig. 11.

High gain combined with low electronic noise can provide energy thresholds clearly below 100 eV. Fig. 12 shows the energy spectra at different gains of the amplifier when no source is used. By applying a cut at the rise time of the signal (which actually provides the depth of the ionized electrons produced in the gas) we can exclude the signal induced by cosmic rays and measure only the Copper energy line of 8 keV (plotted with dashed line in Fig. 12). Then, by increasing the gain of the amplifier 20 times (from Gain = 10 to Gain = 200), we keep in the ADC acceptance only the ultra low energy region (plotted with full line in Fig. 12). The value of 1000 ADC corresponds to ~ 150 eV and the peak at ~ 50 eV is compatible with single electrons (see section 2). Thus, the detection threshold of the Spherical Proportional Counter is as low as 25 eV. From ~ 750 eV to 1200 eV the energy spectrum looks flat (Fig. 12 – full line). At energies lower than ~ 750 eV, an increase of the slope of the energy spectrum is observed and is followed by a steer rise at very low energies. The exponential shape of the spectrum has to be confirmed by underground measurements in the LSM laboratory, which are on-going.

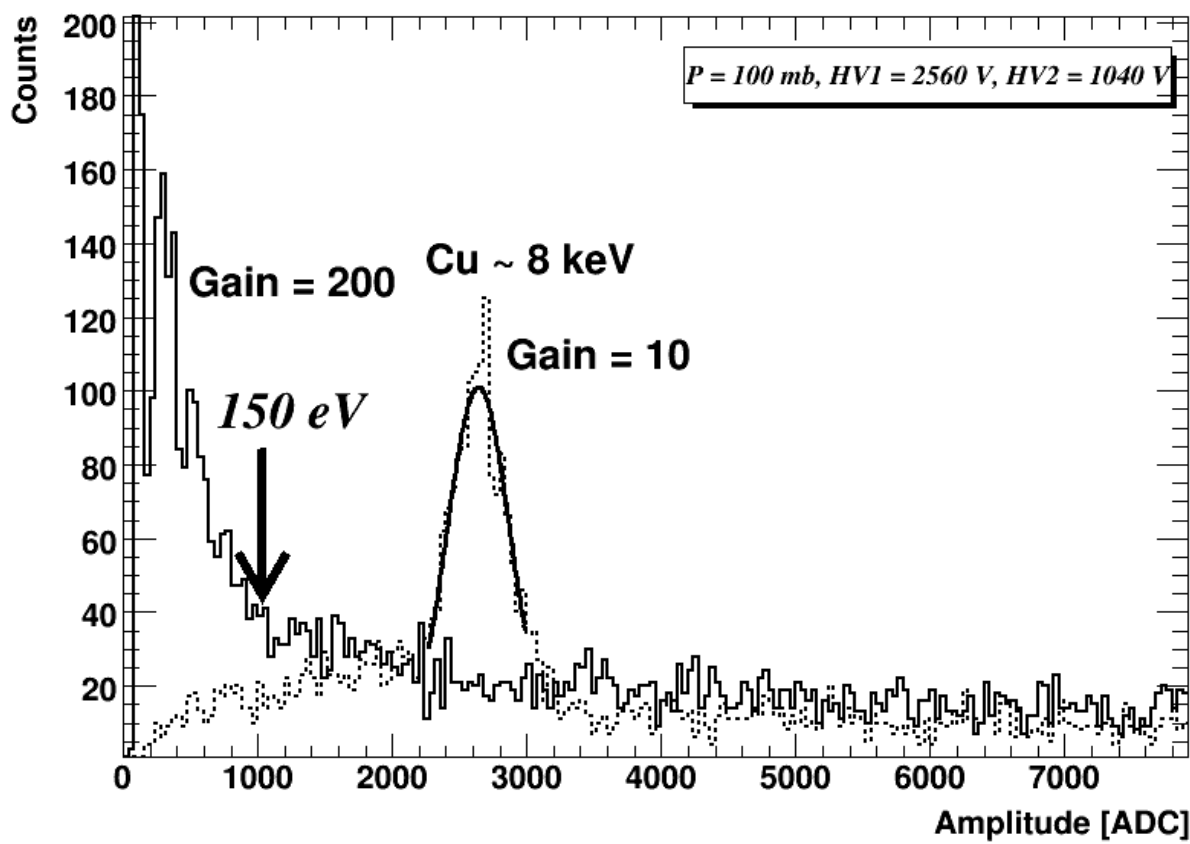

Figure 12: The energy spectrum at Gain = 10 with the Copper peak of 8 keV (dashed line) and at Gain = 200 where the single electron peak at ~ 50 eV is clearly visible (full line).

## 2. Results using a UV flash lamp

A UV window made by $MgF_2$ crystal was installed in one of the sphere openings. A hydrogen relaxation flash lamp has been used to produce UV photons in the far ultra violet range. The discharge is also producing a fast signal which is observed through a capacitor and after adequate attenuation is serving as a fast trigger. Photons are crossing the UV entrance window and hitting copper of the internal vessel producing photoelectrons by photoelectric effect. Electrons are extracting and drifting through the radial field to the central ball where they are amplified and collected. The time difference between the trigger and the later signal is measuring the total drift time. The energy calibration has been done using the 22 keV photon $X$ decay of $^{109}Cd$ source and the 8 keV Cu fluorescence.

The UV light flux of the lamp was adjusted by introducing adequate light attenuators down to the single photo-electron extraction on the Copper. Each sample of data were acquired without trigger during the same running time. The subtraction of the amplitude spectrum of the sample lamp turn-off from those of lamp turn-on suppresses the huge background due to cosmic rays. The gas used was a mixture of Ne with 7% $CH_4$ at a pressure of 100 mbar.

Results are shown on Fig. 13 without light attenuator. The amplitude spectrum of the lamp signal is fitted with a Gaussian function. The mean energy value divided by 36 eV, the photon energy for ionization of gas, gives the mean value, 32.8 of photoelectrons, extracted by each flash of the lamp.

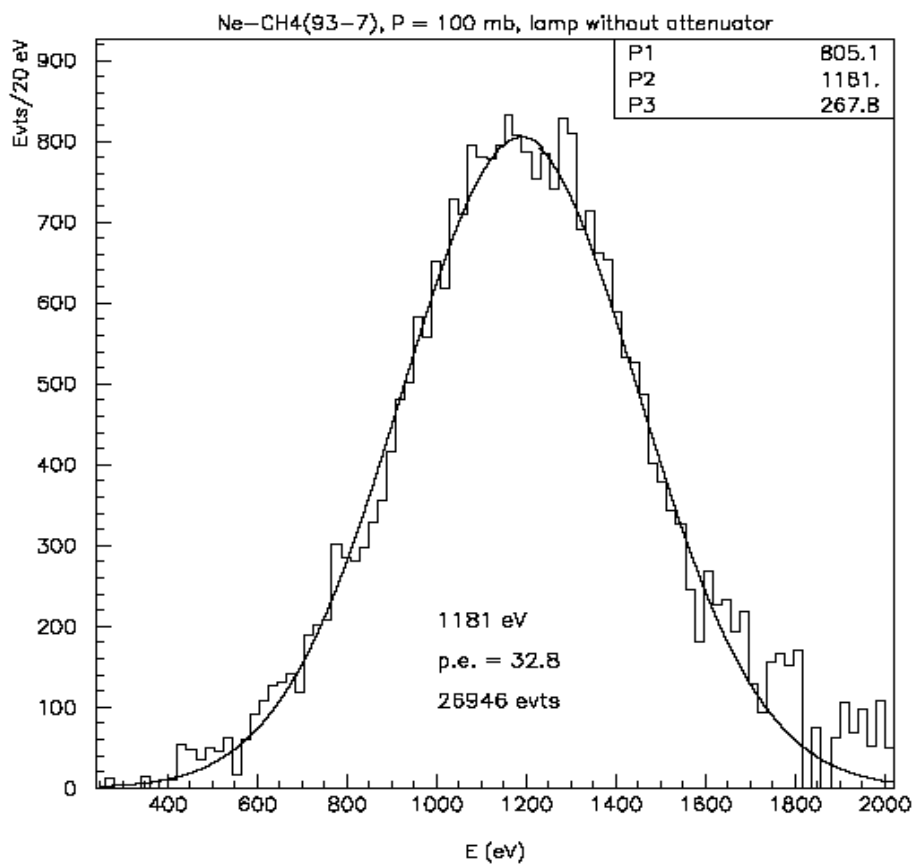

Figure 13: Energy spectrum due to photoelectron extraction from the copper by the UV lamp without attenuation.

Lamp light was then gradually attenuated by 1,2, up to 5 identical attenuators. Each distribution is well fitted by a Polya function with an offset. With less than 3 attenuators, the number of events is stable, which means for those sample the all flashes are detected with a mean frequency close to 45 Hz (27000 events divided by 600 s the run time). In fact the probability not to detect the flash is very low, 2.5% with 2 attenuators, as expected by the Poisson distribution with the mean value given by the fit. With more attenuators the mean frequency is decreasing because the probability of producing zero photoelectrons per flash is increasing exponentially (Fig. 14 - top). The number of photoelectrons per flash given by the fit decreases asymptotically to one as expected. The mean number of extracted electron per flash decrease exponentially with a reduction factor close to 3 as expected (Fig. 14 - bottom).

An example of a good fit using Polya distribution is the case of 3 attenuators giving a mean value of 1.7 photoelectrons per flash as shown in Fig. 15.

The loss of events detected is in good agreement with the number expected by a Poisson distribution, which means an efficiency as large as 80 % to detect the single photoelectrons. Fig. 16 shows the various distributions for all used attenuators. We observe a decrease of the mean value of each distribution up to about 3 attenuators. After that the mean value is not decreasing any more assuring we are at the single electron level.

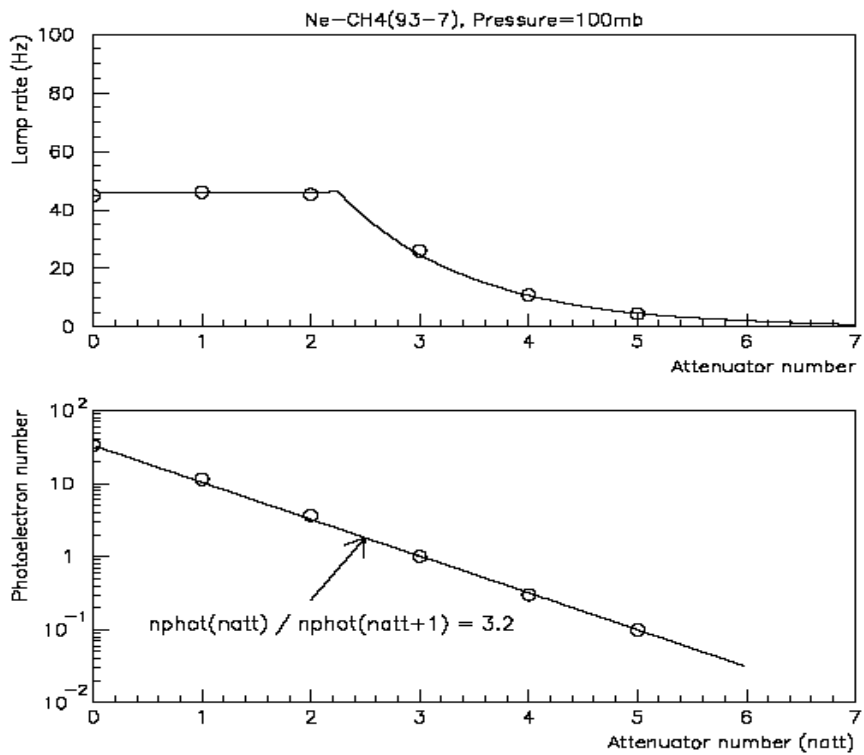

Figure 14: The frequency (top) of detected lamp flashes and the mean number of photoelectrons (bottom) as a function of the number of attenuators.

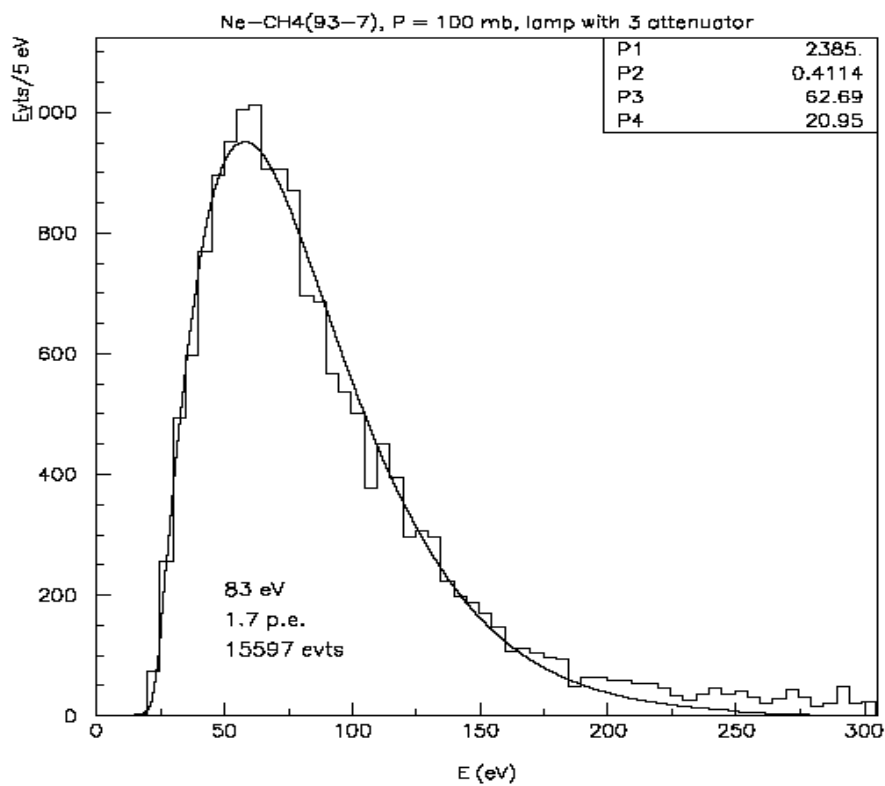

Figure 15: Energy distribution in the case of 3 attenuators.

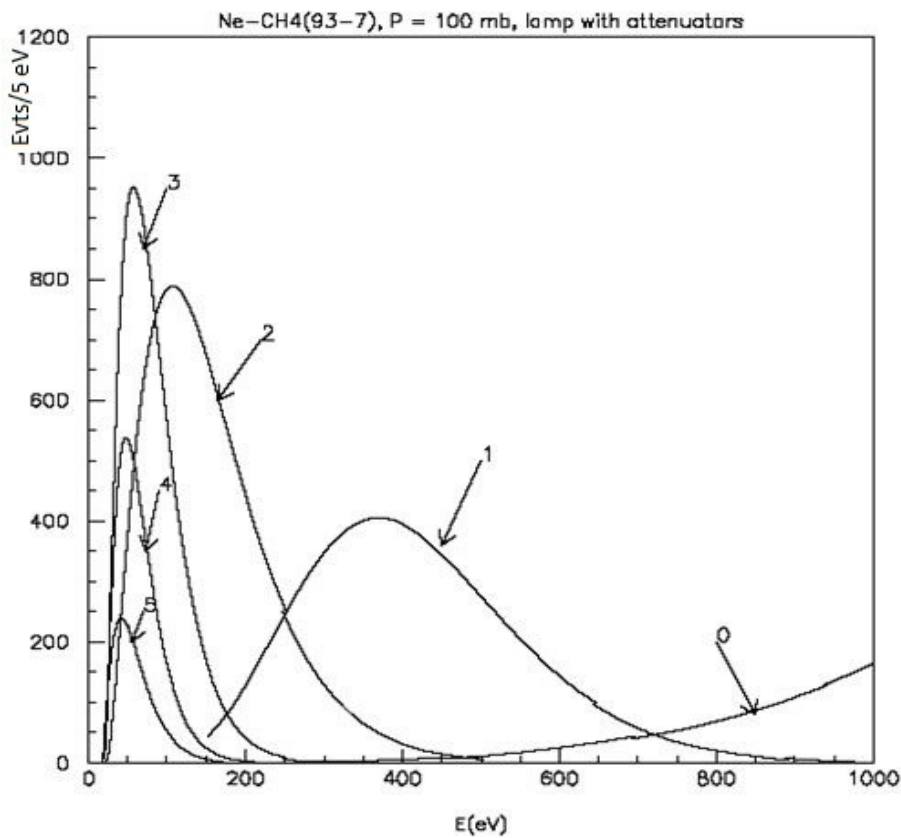

Figure 16: Fitted energy spectrum for several attenuators.

**IV. Future developments and applications**

The counter can be applied for low-background experiments, which require a low energy threshold. An example is the detection of the coherent neutrino-nucleus scattering producing sub keV ion recoils. A spherical detector of radius 4 m and employing Xe gas at a pressure of 10 Atm will detect about 1000 events for a typical supernova explosion at 10 kpc. A world wide network of several such simple, stable and low cost supernova detectors is proposed.

All results that are shown here were taken using the spherical detector of 1.3 m diameter that is made by raw materials. In the underground area another similar detector of the same diameter has been installed in the LSM laboratory in Modane (Frejus) under 1700 m of rock (4800 meters water equivalent) providing protection from cosmic rays. Comparisons between measurements in the sub-keV energy region taken with both detectors are carried out in order to understand the background level and optimize the detector in terms of sensitivity and noise background. The development of a new spherical prototype 70 cm in diameter made of low radioactivity materials is in progress and will be installed in the LSM laboratory with a protected lead wall to eliminate the gamma background and provide measurements of the neutron energy spectrum at the underground installation.

## V. Conclusions

We have developed a new detector with large mass and low sub-keV energy threshold. This is a new type of radiation detector based on the radial geometry with spherical proportional amplification read-out that combines simplicity, robustness and low cost. Several applications are open arising from low energy neutrino physics, dark matter search, Supernova detection to neutron background measurement.

## VI. Acknowledgment

We would like to thank Dr. T. Papaevangelou for exciting discussions.